\documentclass[12pt]{article}
\input epsf
\def\bc{\begin{center}}
\def\ec{\end{center}}
\def\beq{\begin{equation}}
\def\eeq{\end{equation}}
\def\bea{\begin{eqnarray}}
\def\eea{\end{eqnarray}}

\def\phid{\phi^{\dagger}}

\def\pd{\partial}
\def\by{\bar{y}}
\title{\bf On the "toy model" in the Reggeon Field Theory}
\author{M.A.Braun$^{a)}$, and G.P.Vacca$^{b)}$  \\
$^{a)}$ S.Peterburg State University, Russia\\
$^{b)}$INFN and Department of Physics, Via Irnerio 46, Bologna, Italy}
\begin{document}
\maketitle

\begin{abstract}

The Reggeon field theory with zero transverse dimensions is studied in the Hamiltonian
formulation for both sub-and supercritical pomeron. Mathematical aspects of the model,
in particular the scalar products in the space of quantum states, are discussed.
Relation to reaction-diffusion processes is derived in absence of pomeron merging.
Numerical calculations for different parameters of the models, $\alpha(0)-1=\mu$ and the
triple pomeron coupling constant $\lambda$, show that the triple
pomeron interaction  always makes amplitudes fall with rapidity irrespective of the
value of the intercept. The smaller  the  values of the ratio $\lambda/\mu$ the
higher are rapidities $y$ at which this fall starts, so that at small values of
$\lambda$ it begins at asymptotically  high rapidities
(for $\lambda/\mu<1/4$ the fall is
noticeable only at $\mu y>100$).
No visible singularity is seen for the critical pomeron.
A perturbative treatment is proposed which may be useful for more realistic models.
\end{abstract}

\section{Introduction}
At high energies the Quantum Chromodynamics tells us that particle interaction may be
mediated by the exchange of hard pomerons, which are non-local entities propagating  according to
the BFKL equation and splitting into two or merging from two to one with the known triple pomeron
verteces. This most complicated picture can be drastically simplified in the quasi-classical
treatment supposedly valid when at least one of the colliding hadrons is a heavy nucleus.
This leaves only tree diagrams which can be summed by comparatively simple equations, which at least
admit numerical solution.
However at present much attention is drawn to the contribution of loops. This is obviously a
problem of a completely different level of complexity, since it
amounts to solving a full-fledged non-local quantum field theory. At present a solution of even some
local quantum field theory has not been known except for the one-dimensional case.
Naturally for the study of loops in QCD an one-dimensional model has been lately chosen as
a starting point.

The one-dimensional ("toy") model for the QCD interaction was proposed and solved by A.H.Mueller
~\cite{AHM}. However much earlier a model similar in structure was studied in the framework of
a local Reggeon field theory (RFT) with a supercritical pomeron and imaginary triple pomeron interaction
in a whole series of papers ~\cite{amati1,alessandrini,jengo,amati2,ciafaloni1,ciafaloni2}
As expected the one-dimensional reduction of the RFT
admitted more or less explicit solution and lead to definite predictions as to the behaviour of
amplitudes both for hA and AB interaction (of course with the understanding of a highly
artificial physical picture, very far from the realistic world). In the course of study many
ingeneous tricks and techniques were employed and the underlying complicated dynamics was revealed.
However some subtle points have remained unclear in our opinion. In particular the nature of
correspondence between the functional and Hamiltonial approaches
and the origin of restrictions on the form of wave function and the range of its variable
have  not been fully exposed.

To elucidate these points is the primary aim of this note. We also try to follow the line
presented in ~\cite{boreskov,bondarenko} to relate the RFT with the
probabilistic parton picture and the so-called reaction-diffusion model
for the case when there is no merging of pomerons (fan diagrams). This gives us a possibility to
find the contribution from the loops with the highest extention in rapidity just by joining two
fans propagating from the projectile and target towards the center.
We additionally develop  a perturbation theory for small $\lambda$ to compute the evolution
by noting the PT-symmetry of the model.
We finally report on some
numerical results on solving the RFT with all loops, which can be obtained quite easily by
integrating the corresponding evolution equations.

Note that we study exclusively a model with the triple pomeron interaction only.
A different model which  additionally includes a quartic pomeron
interaction with a particular fine-tuned value of the coupling constant
has lately received enough attention in literature
~\cite{boreskov, bondarenko, levin1, levin2}. In our opinion the quartic
interaction has no immediate counterpart in the realistic pomeron theory
with a large number of colours $N_c$, where it is damped by factor $1/N_c^2$.
\section{Functional integral formulation}
The toy model of Pomeron interaction with a zero-dimension
transverse space can be defined by a generating functional
\beq
Z=\int D\phi D\phid e^{-S},
\label{funint}
\eeq
where
\beq
S=\int dy {\cal L},\ \
\eeq
and
\beq
{\cal L}=\frac{1}{2}(\phid\phi_y-\phid_y\phi)-\mu\phid\phi
+i\lambda\phid\phi(\phid+\phi)-i(j_0\,\phid+j_Y\,\phi).
\label{lagrangian}
\eeq
Here $\mu$ is the pomeron intercept ($\alpha(0)-1$). For the
supercritical pomeron $\mu>0$. Triple pomeron coupling constant
$\lambda$ is also positive.
The sources $j_0=\delta(y)g_1$ amd $j_Y=\delta(y-Y)g_2$ represent the
interaction with the
target and projectile respectively. In the real world
$g_1=g_1(b)=AgT_A(b)$ and $g_2=g_2(b)=BgT_B(b)$ where $g$
is the pomeron-nucleon coupling constant, $A$ and $B$ are atomic numbers
of the colliding nuclei and $T_A$ and $T_B$ their profile functions.

The functional integral (\ref{funint}) converges for $\mu<0$ provided the two
field variables $\phi$ and $\phid$ are complex conjugate to one another.
In fact putting $\phi=\phi_1+i\phi_2$ and $\phid=\phi_1-i\phi_2$, with
both
$\phi_{1,2}$ real, we find that all terms in the Lagrangian (\ref{lagrangian}) are pure
imaginary except for the mass term proportional to
\[ \phid\phi=\phi_1^2+\phi_2^2\geq 0.\]
So the integrand in (\ref{funint}) contains a factor
$\exp\Big(\mu\int dy (\phi_1^2+\phi_2^2)\Big)$,
which guarantees convergence if $\mu<0$ and the integration over
the fields $\phi_1$ and $\phi_2$ goes over the real axis. For $\mu>0$
the integral does not exist.
Note that the integral does not exist for any sign of $\mu$ either,
if one changes the
integration contour in $\phi_1$ and $\phi_2$ and so over $\phi$ and
$\phid$. This means that it is impossible to pass to new fields
$u=i\phid$ and $v=i\phi$ and take $u$ and $v$ real directly in the
functional integral as in ~\cite{alessandrini},
 since it requires unlawful rotation of the
integration contour in the integration over $\phi_1$.

Obviously the functional formulation can serve only to define the model
for the subcritical pomeron with $\mu<0$. To define the theory for
positive values of $\mu$ one has to recur to analytic continuation in
$\mu$. Note that
the properties of the functional integral make one think that there is a
singularity
at $\mu=0$. In a fully dimensional RFT it was conjectured that this singularity
was related to a phase transition. However already in ~\cite{ciafaloni1} it
was argued
that there was no phase transition in the zero-dimensional world and the
amplitudes were analytic in $\mu$ on the whole real axis.
As we shall see
by direct numerical calculation, there is indeed  no singularity at
$\mu=0$ in scattering amplitides.

In fact the functional formulation is only supported by the fact that it reproduces
the perturbative diagrams for the pomeron propagation and interaction.
In the following section we introduce an alternative, Hamiltonian formalism, which
gives rise to the same perturbative diagrams (such an approach as a starting point
was briefly mentioned in ~\cite{jengo}).
However in contrast to the
functional approach it does not involve any limitation on the value or sign
of the intercept $\mu$. Since for $\mu<0$ the perturbative series seems to be
convergent,
the two formulations are  completely equivalent for the subcritical pomeron.
Therefore the Hamiltonian formulation gives the desired analytic
continuation of the model to positive values of $\mu$.

\section{Hamiltonian formulation}
\subsection{Basic definitions}
Forgetting the functional integral (\ref{funint}) altogether, we now start from
a quasi-Schroedinger equation in rapidity:
\beq
\frac{d\Psi(y)}{dy}=-H\Psi(y)
\eeq
and postulate the form of the Hamiltonian $H$
\beq
H=-\mu \phid\phi +i\lambda \phid(\phi+\phid)\phi
\label{firstH}
\eeq
as a function of two
Hermithean conjugate operators
$\phi$ and $\phid$
with the commutator
\beq
[\phi,\phid]=1.
\label{commrel}
\eeq
The Hamiltonian is obviously non-Hermithean. We standardly split it
into a free and interaction parts:
\beq
H=H_0+H_I,\ \ H_0=-\mu \phid\phi,\ \ H_I= i\lambda \phid(\phi+\phid)\phi.
\label{Hdecomp}
\eeq

Next step is to define physical observables in this picture.
In accordance with the commutation relation (6\ref{commrel}) we define the vacuum state
$\Psi_0$, normalized to unity, by the condition
\beq
\phi\Psi_0=0.
\eeq
All other states will be built from $\Psi_0$ by application of some number
of operators $\phid$. We
postulate that the transition amplitude from the initial state
$\Psi_i$  at rapidity $y=0$ to the final state $\Psi_f$
at rapidity $y$ is given by
\beq
iA_{fi}=\langle \Psi_f|\Psi_i(y)\rangle=\langle
\Psi_f|e^{-Hy}|\Psi_i\rangle.
\eeq
Here
\beq
\Psi_i(y)=e^{-Hy}\Psi_i
\eeq
is the initial state evolved to rapidity $y$.
The amplitude $A_{fi}$ is imaginary positive so that the matrix element on
the right-hand side of (9) is negative.

It will be convenient to define the initial and final states in terms of some operators
acting on the vacuum state:
\beq
\Psi_i=F_i(\phid)\Psi_0,\ \ \Psi_f=F_f(\phid)\Psi_0.
\eeq
This allows to rewrite the amplitude as a vacuum average
\beq
iA_{fi}=\langle F_{f}^*(\phi)e^{-Hy}F_i(\phid)\rangle,
\eeq
where we do not explicitly write the vacuum state $\Psi_0$ in which the
average is taken.

\subsection{Perturbative expansion}
We are going to prove that  the perturbative expansion of this theory
leads to the same
Feynman diagrams as from the functional approach. To this end we present
\beq
iA_{fi}=\langle
e^{H_0y}F_{f}^*(\phi)e^{-H_0y}e^{H_0y}e^{-Hy}F_i(\phid)\rangle.
\eeq
From (\ref{commrel}) it follows that
\beq
\phi(y)\equiv e^{H_0y}\phi e^{-H_0y}=e^{\mu y}\phi.
\label{phievol}
\eeq
The operator
\beq
U(y)\equiv e^{H_0y}e^{-Hy}
\eeq
satisfies the equation
\beq
\frac{dU(y)}{dy}=e^{H_0y}(H_0-H)e^{-H(y)}=-e^{H_0y}H_ie^{-H_0y}U(y)
=H_I(y)U(y),
\label{inteq}
\eeq
where $H_I(y)$ is the interaction in the interaction representation:
\beq
H_I(y)=e^{H_0y}H_I(\phi,\phid)e^{-H_0y}=H_I(\phi(y),\phid(y)),
\eeq
with the rapidity dependent operators (14) and
\beq
\phid(y)\equiv e^{H_0y}\phid e^{-H_0y}=e^{-\mu y}\phid.
\label{phidevol}
\eeq
Note that at $y>0$ the operators $\phi(y)$ and $\phid(y)$ are no more
Hermithean conjugate to each other, since the normal time
is changed to the imaginary one. However the similarity transformations
(\ref{phievol}) and (\ref{phidevol})
do not change the commutation relations.
The solution of (\ref{inteq}) with the obvious initial condition $U(0)=1$ is
standard
\beq
U(y)=T_y\exp\Big\{-\int_0^y dy'H_i(y')\Big\},
\eeq
where $T_y$ means ordering the operators from left to right according to
decreasing $y$'s. So we get the expression for the amplitude as
\beq
iA_{fi}=\langle F^*_f(e^{\mu y}\phi)T_y\exp\Big\{-\int_0^y
dy'H_i(y')\Big\}
F_i(\phid)\rangle.
\eeq

Expansion of the $T_y\exp$ gives rise to  standard Feynman diagrams.
Verteces for pomeron splitting and merging will be given by the expected
factors
$i\lambda$. The pomeron propagator will be given by
\beq
P(y_1-y_2)=\langle T_y\Big\{\phi(y_1),\phid(y_2)\Big\}\rangle=
e^{\mu(y_1-y_2)}\theta(y_1-y_2),
\eeq
which is the correct pomeron propagator in our toy model.
So indeed the theory defined by this Hamiltonial formulation gives
rise to  standard Feynman diagrams for the RFT.

Note that the Hamiltonian formulation is based on the evolution equation
in rapidity and does not require the intercept $\mu$ to have a definite sign.
It looks equally good both for positive and negative values of $\mu$.
For negative $\mu$ it produces a perturbative expansion which is
identical to that in the functional approach. So it gives the desired
analytic continuation of the functional approach to the region of
positive $\mu$, where the latter looses sense.

\subsection{Passing to a real Hamiltonian}
It is perfectly possible to continue with the originally
defined Schroedinger operators $\phi$ and $\phid$ which are
Hermithean conjugate to each other and satisfy the commutation
relation (\ref{commrel}). However it is convenient to pass to new operators
$u$ and $v$ in terms of which all ingredients of the theory
become real. Following the old papers (starting from ~\cite{amati1}) we define
\beq
u=i\phid,\ \ v=i\phi.
\eeq
The new fields are anti-Hermithean to each other
\beq
u^\dagger=-v,\ \ v^\dagger=-u
\eeq
and satisfy the commutation relation
\beq
[v,u]=-1.
\label{commrel2}
\eeq
However $v\Psi_0=0$ and the states are generated by action of $u$.
So $v$ is the annihilation and $v$ is the creation operator with
abnormal commutation relation (\ref{commrel2}).

In terms of new fields the Hamiltonian becomes real
\beq
H=\mu u v-\lambda u(u+v)v.
\label{Huv}
\eeq
It is important that also the operators creating the initial and final
states become real. It is normally assumed that for the scattering
of two nuclei the initial and final wave functions  should correspond to the
eikonal picture. In terms of creation operator $u$ this leads to
\beq
\Psi_{i(f)}=F_{i(f)}(u)\Psi_0,\ \ F_{i(f)}(u)=1-e^{-g_{i(f)}u}.
\eeq
So the expression for the amplitude is rewritten in terms of
real quantities:
\[
iA_{fi}=\langle F_f(u)\Psi_0|e^{-Hy}|F_i(u)\Psi_0\rangle=
\langle\Psi_0|F^*_f(-v)e^{-Hy}F_i(u)\Psi_0\rangle\]\beq=
\langle \Big(1-e^{-g_fv}\Big)e^{-Hy}\Big(1-e^{-g_iu}\Big)\rangle.
\eeq
Since $H\Psi_0=0$ the term independent of $g_i$ and $g_f$ vanishes, so
that
we can also write
\beq
iA_{fi}=
-\langle e^{-g_fv}e^{-Hy}\Big(1-e^{-g_iu}\Big)\rangle=
-\langle\Psi_0|e^{-g_fv}F_i(y,u)\Psi_0\rangle,
\label{ampli1}
\eeq
where $F_i(y,u)$ is the operator which creates the evolved initial state.
Since
\beq
\Psi_i(y)=e^{-Hy}\Psi_i=e^{-Hy}F_i(u)\Psi_0\equiv F_i(y,u)\Psi_0,
\eeq
we have
\beq
\frac{\pd F_i(y,u)}{\pd y}=-H(u,v)F_i(y,u)
\label{evol1}
\eeq
with an initial condition
\beq
F_i(0,u)=1-e^{-g_iu}.
\label{initF}
\eeq
The commutation relation (\ref{commrel2}) allows to represent
\beq
v=-\frac{\pd}{\pd u}
\label{vrepr}
\eeq
and then (\ref{ampli1}) implies that to find the amplitude one has to substitute
$u$ by $g_f$ in $F_i(y,u)$
\beq
iA_{fi}=-F_i(y,g_f).
\eeq

This gives a practical recipe for the calculation of the amplitude.
One has to solve Eq. (\ref{evol1}) with the Hamiltonian (\ref{Huv}) in which $v$ is
represented according to (\ref{vrepr}):
\beq
H=-\mu u \frac{\pd}{\pd u}+\lambda u^2\frac{\pd}{\pd u}-
\lambda u\frac{\pd^2}{\pd u^2}
\label{Huvdiff}
\eeq
and the initial condition (\ref{initF}). After the  solution is found,
one has to substitute
$u$ by $g_f$ in it.

\subsection{Target-projectile symmetry}
Taking the complex conjugate  of (\ref{ampli1}) we find
\beq
-iA^*_{fi}=
\langle \Big(1-e^{g_iv}\Big)e^{-H^{\dagger}y}\Big(1-e^{g_fu}\Big)\rangle
=iA_{if}(\lambda\to -\lambda,g_{i(f)}\to -g_{f(i)}).
\eeq
Having in mind that the amplitude is pure imaginary,
we see that interchanging the target and projectile leads to changes of
sign in
the triple pomeron coupling and the couplings to the external particles.
On the other hand changing $u\to -u$ in Eq. (\ref{evol1}) with the Hamiltonian
(\ref{Huvdiff}) and the initial condition
(\ref{initF}) and also changing signs of $\lambda$ and $g_i$  obviously does not
change the solution:
\beq
F_i(y,-u,-g_i,-\lambda)=F_i(y,u,g_i,\lambda),
\eeq
from which we find
\beq
F_i(y,-g_f,-g_i,-\lambda)=F_i(y,g_f,g_i,\lambda).
\eeq
So  the interchange of the target and projectile
does not change the amplitude.

\subsection{Scalar products}
As follows from our derivation the exact mathematical realization of the
scalar product in the space of wave functions $\Psi$ is irrelevant for
the resulting formulas. The representation  of the operator $v$ as a
minus derivative in $u$ serves only to express its algebraic action on
the operator $u$ in accordance with the commutation relation.
It does not require to introduce a representation for the wave function $\Psi$
in which $u$ is represented as multiplication by a complex number and
correspondingly the scalar product is defined by means of  integration
over the whole complex plane $u$ with a weight factor $e^{-uu^*}$
aas in ~\cite{ciafaloni1}.
One may instead represent operators $u$ and $v$ in terms of the standard
Hermithean operators $p$ and $q$ as
\beq
u=\frac{i}{\sqrt{2}}(q-ip),\ \ v=\frac{i}{\sqrt{2}}(q+ip)
\eeq
and define a scalar product as
\beq
\langle\Psi_1|\Psi_2\rangle=\int dq\Psi^*_1(q)\Psi_2(q)
\eeq
with $p=-i\pd/\pd q$. The vacuum state $\Psi_0$ will be given by the
oscillator ground state in this representation.

However this is not the end of the story. New aspects  arise in the
process of solution of the Eq. (\ref{evol1}) for $F_i$. The equation itself is
regular everywhere except
possibly at $u=0$, where however it is also regular having behaviour as a
constant or $u$. We are obviously interested in the last behaviour since
from the equation it follows that
$F_i(y,u=0)=0$.
So in principle our solution turns out to be regular in the whole
complex $u$-plane. Formally this means that one may choose any continuous
interval in this plane and evolve the initial function given in this interval up
to the desired values of rapidity. The immediate problem is what interval one has
to choose. The final expression for amplitude requires that the value of $g_f$ has
to belong to this interval. In particular for positive $g_f$ the interval should
include at least a part of the positive axis.

Numerical calculations show that it is indeed possible to find a solution to the
evolution equation provided the initial interval is restricted  to {\it only}  positive
values of $u$ (including the desired value $u=g_f$).

To understand this phenomenon one has to recall some results from the earlier
studies.
It was found in ~\cite{jengo} that there exists a transformation of both the
function and variable
which  converts the non-Hermithean Hamiltonian (\ref{Huvdiff}) into a
Hermithean one.
However it is only possible for values of $u$ covering the positive axis rather than
the whole real axis where the equation should hold.
Namely one introduces  variable $x$ by
\beq
u=x^2>0
\label{upos}
\eeq
and makes a similarity transformation of the Hamiltonian
\beq
\tilde{H}=W^{-1}HW, \ \
W(x)=\sqrt{x}e^{\frac{1}{4}x^4-\frac{\mu}{2\lambda}x^2}
\label{iengoform}
\eeq
to find the transformed Hamiltonian  as
\beq
\tilde{H}=\frac{\lambda}{4}\Big(-\frac{\pd^2}{\pd x^2}+\frac{3}{4x^2}+x^2(x^2-\rho)^2-2x^2\Big)
\label{Hiengoform}
\eeq
where we standardly denote $\rho=\mu/\lambda$.
The Hamiltonian $\tilde{H}$ is a well defined Hamiltonian on  the whole real axis with
a singularity at $x=0$. The interaction potential infinitely grows at $|x|\to \infty$.
So restricting to $x>0$ and
requiring  eigenfunctions to vanish at the origin and at $x\to\infty$ one finds a
discrete spectrum, starting from a certain ground state $E_0$ to infinity.
The transformed initial function
\beq
\tilde{F}_i(y=0,x)=\frac{1}{\sqrt{x}}e^{-\frac{1}{4}x^4+\frac{\mu}{2\lambda}x^2}
\Big(1-e^{-g_ix^2}\Big)
\eeq
is well behaved both at $x=0$ and $x=\infty$ and at $0<x<\infty$ can be expanded in the
complete set of eigenfunctions $\tilde{F}_n$ of $\tilde{H}$:
\beq
\tilde{F}_i(y=0,x)=\sum_{n=0}\langle
\tilde{F}_i(y=0,x)|\tilde{F}_n\rangle \tilde{F}_n(x).
\label{initF2}
\eeq
Here a new scalar product has been introduced for the transformed functions:
\beq
\langle\tilde{F}_1|\tilde{F}_2\rangle=\int_0^\infty dx\tilde{F}^*_1(x)\tilde{F}_2(x).
\eeq
Evolution immediately gives
\beq
\tilde{F}_i(y,x)=\sum_{n=0}\langle
\tilde{F}_i(y=0,x)|\tilde{F}_n\rangle e^{-E_ny}\tilde{F}_n(x)
\eeq
and applying the operator $W$ one finds the solution as
\beq
F(y,u)=\sum_{n=0}\langle\tilde{F}_i(y=0,x)|\tilde{F}_n\rangle
e^{-E_ny}W(x)\tilde{F}_n(x),
\ \ u=x^2.
\eeq

Note that we could just as well start from negative values of $u$ and define
instead of (\ref{upos})
\beq
u=-x^2,\ \ x>0
\eeq
Doing the same transformation of the Hamiltonian we then obtain (\ref{Hiengoform}) with the
opposite
overall sign and an opposite sign of $\rho$,
that is the transformed Hamiltonian is now $-\tilde{H}(\rho\to -\rho)$. All eigenvalues now
are negative and go down to $-\infty$.
If we take the expansion of the initial function analogous to (\ref{initF2}) and try to
evolve it we obtain a sum
with infinitely growing exponentials in $y$. Obviously such evolution has no sense.
This explains why with the positive or negative signs of $\lambda$ and
$g$'s one can start evolution
only from correspondingly positive or negative values of $u$, but never
from intervals including both.

Also note that considered as a function of $u$, eigenfunctions are analytic
in the whole $u$-plane. Starting from, say, $u>0$ their values at $u<0$ can be obtained by
analytic continuation to pure imaguinary values of variable $x$. Obviously eigenfunctions
should exponentially grow as $u\to -\infty$ (otherwise we shall find positive eigenvalues
of the continued Hamiltonian $-\tilde{H}(\rho\to -\rho)$).
Similarly starting from $u<0$ and passing to
positive $u$ one finds that eigenfunctions exponentially grow at large positive $u$.
So considering the initial Hamiltonian $H$ on the whole real $u$ axis we find that its
spectrum goes from $-\infty$ to $+\infty$ with eigenfunctions which vanish only either
at large positive $u$ for positive eigenvalues or at large negative $u$ for negative
eigenvalues and exponentially grow otherwise. This prevents introduction of the
Bargmann scalar product with integration over the whole complex $u$-plane.

Finally let us investigate the qualitative behaviour of the system in the
large $\lambda/\mu=1/\rho$ limit, which may be of interest in the generalization
to more dimensions, where the effective $\lambda$ is (infinitely) large
(see \cite{amati2}).
For this purpose the representation given in Eq. (\ref{Hiengoform}) is
convenient, since it admits  perturbative expansion in $\rho$.
In the limit $\rho=0$ one has
\beq
\tilde{H}=
\frac{\lambda}{4}\Big(-\frac{\pd^2}{\pd x^2}+\frac{3}{4x^2}+x^6-2x^2\Big)
\label{iengoform0}
\eeq
which means that the amplitude will behave as
$e^{- y E_0}$, where $E_0$ is the ground state energy of $\tilde{H}$.

One can perform a semiclassical estimation of the ground state of the
Hamiltonian in Eq. (\ref{iengoform0}) by imposing the condition
\beq
\int_{x_1(4E/\lambda)}^{x_2(4E/\lambda)} dx
  \sqrt{4E/\lambda-V(x)}=\left(n+\frac{1}{2}\right)\pi,
\label{semiclassical}
\eeq
where $V(x)=\frac{3}{4x^2}+x^6-2x^2$ and $x_i$ are the positive real solutions
of the equation $V(x)=4 E/\lambda$.
For $n=0$ one obtains $E_0 \simeq 3.7 \, \lambda/4 >0$, which  is in a
very good agreement  with the results of  numerical
evolution for $\lambda/\mu \ge 3$, which indeed shows
a decay of the amplitude
as $e^{-y E_0}$ after a short initial period of evolution.

Note that during this
initial period one observes a rise of the amplitude, which may seem strange in view
of the fact that all eigenvalues of the transformed Hamiltonian (\ref{iengoform})
are in fact positive. This rise is explained by the interference of different
contributions which may enter with different signs due to the non-unitarity
of the similarity transformation (\ref{iengoform}) from the initial non-Hermithean Hamiltonian.

 \section{No pomeron fusion: fan diagrams}
\subsection{Solution}
A drastic simplification occurs if in the Hamiltonian one drops the term corresponding to
fusion of pomerons ('fan case'):
\beq
H\to H_{fan}=\mu uv-\lambda u^2v=-(\mu u-\lambda u^2)\frac{\pd}{\pd u}.
\label{Hfan}
\eeq
Then, as has been known since ~\cite{schwimmer}, the solution can be
immediately obtained in an explicit form.
Indeed, following ~\cite{boreskov}, define a new variable $x$ by
\beq
 dz=\frac{du}{\mu u-\lambda u^2},
\eeq
which gives (up to an irrelevant  constant)
\beq
z=\frac{1}{\mu}\ln\frac{u}{u-\rho},\ \ \rho=\frac{\mu}{\lambda}.
\eeq
The inverse relation is
\beq u=\frac{\rho}{1-e^{-\mu z}}.
\eeq
Our evolution equation for $F$ takes the form
\beq
\Big(\frac{\pd}{\pd y}-\frac{\pd}{\pd z}\Big)F(y,z)=0
\eeq
with a general solution
\beq
F(y,z)=f(y+z)
\label{yshift}
\eeq
where $f$ is an arbitrary function. It is to be found from the initial
condition:
\beq
F(0,z)=f(z)=1-e^{-g_iu}=1-e^{-g_i\frac{\rho}{1-e^{-\mu z}}}.
\eeq
From this we obtain~\cite{boreskov}
\beq
F_{fan}(y,u)=1-\exp\Big(-g_i\frac{\rho}{1-e^{-\mu (y+z)}}\Big)
=1-\exp \Big[-\frac{g_iue^{\mu y}}{1+\frac{u}{\rho}\Big(e^{\mu
y}-1\Big)}\Big].
\label{fansol1}
\eeq
This solution is evidently non-symmetric in the projectile and target.
In practice one considers such an approximation in the study of hA
scattering. Then it represents the sum of fan diagrams propagating from
the projectile hadron towards the target nucleus. This situation
corresponds to the lowest order in powers of $g_i$~\cite{schwimmer}:
\beq
F^{hA}_{fan}=\frac{g_iue^{\mu y}}{1+\frac{u}{\rho}\Big(e^{\mu y}-1\Big)}
\label{fansolhA}
\eeq
We recall that the amplitude is obtained from (\ref{fansol1}) or (\ref{fansolhA}) by putting
$u=g_f$

We note that the solution (\ref{fansolhA}) is not analytic in the whole complex $u$
plane: it evidently has a pole at some negative value of $u$, which
tends to zero from the negative side as $y\to\infty$. So at least for this
solution one cannot formulate a scalar product of the Bargmann type with
integration over the whole complex $u$ plane. As we have stressed,
the existence of such a scalar product is not at all needed.

In fact the Hamiltonian $H_{fan}$ has the same spectrum as the free
Hamiltonian $H_0$ (Eq. (\ref{Hdecomp})). The two Hamiltonians are related by a
similarity transformation
\beq
H_{fan}=VH_0V^{-1},\ \ V=e^{-u^2v/\rho}.
\label{similfan}
\eeq
It follows that eigenvalues of $H_{fan}$ are all non-positive:
$E_n=-\mu n$ with $n=0,1,2,...$. However the amplitudes (\ref{fansol1}) and (\ref{fansolhA})
do not grow but rather tend to a constant at large $y$. As seen from the
structure of expressions (\ref{fansol1}) and (\ref{fansolhA}) this is a result of interference of
contributions with different $E_n$, which separately grow at high
$y$.
One can reproduce the result of the fan evolution using the form in
Eq. (\ref{similfan}) by noting that $e^{-y H_0}$ is an operator which shifts
the variable $\ln u$ by $\mu$ while $V$ corresponds to a shift in the variable
$1/u$ by $1/\rho$.

The fan evolution is to be compared with the situation with the total Hamiltonian
$H$, whose eigenvalues are all non-negative for the branch with positive
$u$. In this case the solutions generally fall at large $y$ and became
nearly constant only at very small $\lambda$ when one of the excited
levels goes down practically to zero due to a particular structure
of the transformed Hamiltonian (\ref{Hiengoform}) ~\cite{jengo}. So the mechanism
for the saturation of the amplitudes at large rapidities is completely
different for the complete Hamiltonian $H$ and its fan part $H_{fan}$.

\subsection{Relation to the reaction-diffusion approach}
The fan case admits a reinterpretation in terms of the so-called
reaction-diffusion processes ~\cite{bondarenko}, which may turn out to be helpful for
applications.
Let us present a solution $F(y,u)$ as a power series in $u$
\beq
F(y,u)=\sum_{n=1}c_n(y)u^n.
\label{powerexp}
\eeq
The evolution equation gives a system of equations for coefficients
$c_n(y)$:
\beq
\frac {dc_n(y)}{dy}=\mu n c_n(y)-\lambda (n-1)c_{n-1}(y).
\eeq

Consider first the case $\mu>0$.
We rescale $y=\bar{y}/\mu$ and present
\beq
c_n(\bar{y})=\frac{1}{n!}(-\rho)^{1-n}\nu_n(\bar{y})
\label{cn}
\eeq
to obtain an equation for $\nu_n$
\beq
\frac{d\nu_n(\bar{y})}{d\bar{y}}= n
\nu_n(\bar{y})+n(n-1)\nu_{n-1}(\bar{y}).
\label{nuevol}
\eeq
This equation is identical with the one which is obtained for
multiple moments of the probability $P_k(\bar{y})$ to have exactly $k$ pomerons
\beq
\nu_n(\bar{y})=\sum_{k=n}P_k(\bar{y})\frac{k!}{(k-n)!}
\eeq
provided the probabilities obey the equation
\beq
\frac{dP_n(\bar{y})}{d\bar{y}}=-nP_n(\bar{y})+(n-1)P_{n-1}(\bar{y}).
\label{probevol}
\eeq
The equations for $P_n$ are such that they conserve the total probabilty
assumed to be equal to unity. Also if the initial probabilities are
non-negative, they
will remain such during the evolution. This gives the justification
for the interpretation of $P_n$ as probabilities.

For further purposes it is convenient to introduce a generating functional for
the probabilities $P_n$
\beq
Z(\bar{y},u)=\sum_nP_n(\bar{y})u^n.
\eeq
In terms of this functional
\beq
P_n=\frac{1}{n!}\Big(\frac{\pd}{\pd u}\Big)^n Z_{\Big|u=0},\ \
\nu_n=\Big(\frac{\pd}{\pd u}\Big)^n Z_{\Big|u=1}.
\eeq
From the system of equation for $P_n$ one easily finds an equation for $Z(\bar{y},u)$
\beq
\frac{\pd Z(\bar{y},u)}{\pd\bar{y}}=(u^2-u)\frac{\pd Z(\bar{y},u)}{\pd u}.
\eeq
Note that this is essentially the same equation as for $F$ with parameters put to unity and
the  opposite sign of the Hamiltonian. So its solution is immediately
obtained from (\ref{yshift}) by
changing sign of $y$
\beq
Z(\bar{y},u)=Z\Big(0,z(u)-\bar{y}\Big),\ \  u=\frac{1}{1-e^z}.
\eeq

For the simple case of hA scattering this gives
\beq
Z^{hA}=\frac{ue^{-\by}}{1+u(e^{-\by}-1)}.
\eeq
Expansion in powers of $u$ gives the probabilities $P_n$ as
\beq
P_n^{hA}=e^{-\by}\Big(1-e^{-\by}\Big)^{n-1}.
\label{PnhA}
\eeq
For a general case we  first construct $Z(y=0)$ as a sum
\beq
Z(0,u)=1+\sum_{n=1}\nu_n(0) \frac{(u-1)^n}{n!}
\eeq
with values of $\nu_n(0)$ known from the initial distribution:
\beq
\nu_n(0)=\rho^{n-1}g_i^n.
\eeq
Subsequent summation and shift $z\to z-\by$ lead to the final result
\beq
Z(y,u)=1-\frac{1}{\rho}+\frac{1}{\rho}
\exp\Big(g_i\rho
\frac{(u-1)e^{\by}}{u-(u-1)e^{\by}}
\Big).
\eeq
The probabilities should be obtained by developing this expression around $u=0$, which is
not easily done in the general form.

Now let us briefly comment on the case of the subcritical pomeron
$\mu=-\epsilon<0$.
We again rescale
\beq
y=y'/\epsilon
\eeq
and instead of (\ref{cn}) put
\beq
c_n(y')=\frac{1}{n!}\rho^{1-n}\nu'_n(y')
\eeq
where now $\rho=-\epsilon/\lambda<0$.
The resulting equations for $\nu'_n$ are
\beq
\frac{d\nu'_n(y')}{d y'}=
-n\nu'_n(y')+n(n-1)\nu'_{n-1}(y').
\eeq
As we see they become similar to the equations for $P_n$
in the case of $\mu>0$ and in fact coincide with them for
the quantities $\tilde{\nu}_n=\nu'_n/n!$.
This means that for negative $\mu$ the relation between the
RFT and reaction-diffusion processes is different from the case of
positive $\mu$. In fact the interpretation of the
amplitude in terms of $P_n$ and $\nu_n$ interchanges: the
rescaled coefficients $c_n$ in the amplitude give directly the
probabilities to find a given number of pomerons and their
multiple moments have to be defined in terms of these quantities
in the way similar to the construction of probabilities
from their moments for $\mu>0$.
\section{Big loops}
One can use the Hamiltonian approach to find the contribution of
loops with the maximal extention in rapidity, similar to the
approach first taken in~\cite{salam} in the fully dimensional
dipole picture. This issue has been also investigated for a different model
with a specific quartic interaction where an exact solution can be built~\cite{levin1}.
To this end we first split the evolution in two parts:
\[
iA_{fi}(y)=\langle F^*_f(-v)e^{-Hy}F_i(u)\rangle\]\beq=
\langle F^*_f(-v)e^{-Hy/2}e^{-Hy/2}F_i(u)\rangle=
\langle \Big(e^{-H^{\dagger}y/2}F_f(u)\Big)^\dagger
e^{-Hy/2} F_i(u)\rangle.
\eeq
Next we neglect fusion of pomerons during the first part of the evolution
and merging of them during the second:
\beq
iA_{fi}(y)=\langle \Big(e^{-H_{fan}^{\dagger}y/2}F_f(u)\Big)^\dagger
e^{-H_{fan}y/2} F_i(u)\rangle.
\label{amplBL}
\eeq
Here $H_{fan}$ is given by Eq.(\ref{Hfan}) and $H_{fan}^{\dagger}$ differs from
it by the sign of $\lambda$.
This approximation takes into account loops which are obtained by joining
at center rapidity the two sets of fan diagrams going  from
the target and projectile.

Both operators inside the vacuum average are known and given by Eqs.
(\ref{fansol1}) or (\ref{fansolhA}) with the understanding that for the solution with
$H_{fan}^{\dagger}$ one has to change the sign of $\lambda$.
To simplify we restrict to the lowest order in both coupling constants
$g_{i(f)}$. Separating $g_ig_f$ we then obtain the pomeron Green function
with loops of the maximal extention.
In this case the right factor in (\ref{amplBL}) is given by
\beq
e^{-H_{fan}y/2}u=\frac{ue^{\mu y/2}}{1+\frac{u}{\rho}
\Big(e^{\mu y/2}-1\Big)}
\eeq
and the right factor is
\beq
\Big(e^{-H_{fan}^{\dagger}y/2}u\Big)^\dagger=
\Big[\frac{ue^{\mu y/2}}{1-\frac{u}{\rho}
\Big(e^{\mu y/2}-1\Big)}\Big]^\dagger=
\frac{-ve^{\mu y/2}}{1+\frac{v}{\rho}
\Big(e^{\mu y/2}-1\Big)}.
\eeq
So the pomeron Green function is
\beq
G(y)=-\langle\frac{ve^{\mu y/2}}{1+\frac{v}{\rho}
\Big(e^{\mu y/2}-1\Big)}\frac{ue^{\mu y/2}}{1+\frac{u}{\rho}
\Big(e^{\mu y/2}-1\Big)}\rangle.
\eeq
Its calculation is trivial.
Denoting
\beq
b=\frac{1}{\rho}\Big(e^{\mu y/2}-1\Big)
\eeq
we present the left operator as
\beq
\frac{ve^{\mu y/2}}{1+\frac{v}{\rho}
\Big(e^{\mu y/2}-1\Big)}=e^{\mu y/2}\frac{v}{1+bv}=
e^{\mu y/2}\frac{1}{b}\Big(1-\frac{1}{b}\frac{1}{v+1/b}\Big).
\eeq
Action of the unit operator gives zero so that we are left with
\beq
G(y)=e^{\mu y}\frac{1}{b^2}\langle \frac{1}{v+1/b}\frac{u}{1+bu}\rangle.
\eeq
Now we present
\beq
\frac{1}{v+1/b}=\int_0^\infty dxe^{-x(v+1/b)}
\eeq
and, since  operator $e^{-xv}$ just substitutes $u$ by $x$ in the vacuum average,
obtain
\beq
G(y)=e^{\mu y}\frac{1}{b^2}\int_0^{\infty}dx e^{-x/b}\frac{x}{1+bx}=
e^{\mu y}\frac{1}{b^2}\Big(1+\frac{1}{b^2}e^{1/b^2}{\rm Ei}\,
(-1/b^2)\Big).
\label{pomBLsol}
\eeq
In the asymptotic limit $y\to\infty$, in this approximation,
the propagator tends to a constant value
\beq
G(y)_{y\to\infty}\simeq \rho^2\frac{e^{\mu y}}{\Big(e^{\mu y/2}-1\Big)^2}
\simeq \rho^2.
\eeq
The loops tamed its initial exponential growth as $e^{\mu y}$ illustrating the
saturation phenomenon in the old Gribov's RFT with the supercritical
pomeron.

Note that the same result can be obtained by developing in the number
of interactions~\cite{salam,Kovchegov}. We find worth to remind to the reader
how one proceeds in this case. In fact we can expand both wave
funcions in powers of $u$ and $v$ respectively as in (\ref{powerexp}).
Obviously only terms with equal numbers of $u$ and $v$ contribute with
\beq
\langle v^nu^n\rangle=(-1)^n n!.
\eeq
So we get
\beq
G(y)=-\sum_{n=1}(-1)^n n!c_n^2(y/2),
\label{Gcn}
\eeq
where from (\ref{fansolhA}) we find
\beq
c_n(y/2)=e^{\mu y/2}(-b)^{n-1}.
\label{cnyhalf}
\eeq
We find a divergent series
\beq
G(y)=-\frac{1}{b^2}e^{\mu y}\sum_{n=1}(-b^2)^n n!
\eeq
which is however  Borel summable. Presenting
\beq
n!=\int_0^{\infty}dt e^{-t} t^n
\eeq
we get
\beq
G(y)=-\frac{1}{b^2}e^{\mu y}\int_0^\infty dt e^{-t}\sum_{n=1}(-b^2t)^n
=e^{\mu y}\int_0^\infty dt e^{-t}\frac{t}{1+b^2t}.
\eeq
This is the same expression (\ref{pomBLsol}) which we obtained before.
So these simple but not very rigorous manipulations give the correct answer.

It is instructive to express the same result in terms of the probabilities
$P_n$. Actually it is a straightforward task. Rescaling the coefficients
$c_n$ in (\ref{Gcn}) we rewrite this expression in terms of multiple probability
moments
$\nu_n$
\beq
G(y)=-\rho^2\sum_{n=1}\Big(-\frac{1}{\rho^2}\Big)^n\nu_n^2(y/2)\frac{1}{n!}.
\label{Gnun}
\eeq
Now we express the multiple moments in terms of the probabilities according to
(\ref{probevol}) to finally obtain
\beq
G(y)=\sum_{k,l=1}P_k(y/2)P_l(y/2)F_{kl},
\eeq
where $F_{kl}$ are coefficients independent of energy
\beq
F_{kl}=\sum_{n=1}n! C_k^nC_l^n\Big(-\frac{1}{\rho^2}\Big)^{n-1}.
\eeq

It is easy to find the asymptotical values of $F_{kl}$ at large $k$ and $l$.
Then we can approximate
\beq
C_k^n\simeq \frac{k^n}{n!},\ \ n<<k\to\infty.
\eeq
Under this approximation
\beq
F_{kl}=-\rho^2\sum_{n=1}\frac{1}{n!}\Big(-\frac{kl}{\rho^2}\Big)^n=
\rho^2\Big(1-e^{-kl/\rho^2}\Big).
\eeq
Then we get from (\ref{Gnun})
\beq
G(y)=\rho^2\Big(1-\sum_{k,l=1}P_k(y/2)P_l(y/2)e^{-kl/\rho^2}\Big)
\eeq
At high $y$ and fixed $n$ the probabilities $P_n(y)$ become independent of
$n$ and equal to $e^{-\mu y}$
(see Eq. (\ref{PnhA})). Then we get
\beq
G(y)=\rho^2\Big(1-e^{-\mu y}\sum_{k,l=1}e^{-kl/\rho^2}\Big),
\eeq
with the correct limiting value $\rho^2$ at $y\to\infty$.
\section{Perturbative analysis for small $\lambda$ and PT-symmetry}

Having in mind possible generalizations to the realistic two-dimensional world,
in this section we study  perturbative treatment of the model at
small values of $\lambda$. As follows from the previous studies
~\cite{jengo,amati2,ciafaloni1}
actually the point $\lambda=0$ corresponds to an essential singularity
in the Hamiltonian spectrum corresponding to the existence of a low-
lying excited state with a positive energy
\beq
\Delta E=\frac{\mu^2}{\sqrt{2\pi} \lambda} e^{-\rho^2/2}.
\label{DeltaE}
\eeq
This states dominates  evolution at very high rapidities.
However at  earlier stages of evolution this unperturbative component
may play a secondary role as compared to purely perturbative contributions.
We also point out in this section some general properties of the   Hamiltonian $H$
given by Eq. (5),which
belongs to a class of non-Hermithean Hamiltonians widely studied in
literature.

Inspecting the Hamiltonian $H$ we can see that even
if it is not Hermithean, it describes a PT-symmetric  system
evolving in the imaginary time. This property has been intensively studied mainly for
anharmonic oscillators \cite{bender}.
Indeed $H$ is invariant under the product of parity and ''time''-reversal
transformations: $\phi \to -\phi$, $\phi^\dagger \to - \phi^\dagger$
and $i \to -i$.

As the transformation to the form (42) has shown, the spectrum of $H$ is real.
This fact  is
true whenever for a
diagonalizable operator $H$ certain conditions are fullfilled \cite{mostafazadeh}.
The most important is that $H$  should be Hermithean with respect to some
positive-definite scalar product $\langle \cdot | \cdot \rangle_\eta =
\langle \cdot | \eta \cdot \rangle$,
constructed using a positive-defined metric operator $\eta=e^{-Q}$
with Hermithean $Q$.
This defines the $\eta$-pseudo-Hermiticity property of $H$, which
may be conveniently written as
\beq
H^\dagger=\eta\, H \,\eta^{-1} = e^{-Q} H \, e^Q.
\label{pseudoH}
\eeq
Considering the set  all positive-definite metric operators, it is
easy to see that the parity operator $P$ belongs to it.
Then one can define a $C$ operator by $C=P^{-1} \eta= \eta^{-1} P$ which
commutes with the Hamiltonian $H$ and the $PT$ symmetry generator.
In this formalism one should therefore introduce as a scalar product the
CPT-inner scalar product
\beq
\langle \Psi_1 |\Psi_2 \rangle_{CPT}=\langle \Psi_1|e^{-Q}\Psi_2\rangle
=\langle e^{-Q/2}\Psi_1|e^{-Q/2}\Psi_2\rangle
\eeq

The $\eta$ operator is in general unique up to symmetries of $H$. It
describes the nature of the physical Hilbert space, the observables being
$\eta$-pseudo-Hermithean operators.
Using a symmetric (second) form of the scalar product (105)
we define
\beq
h=e^{-Q/2} H \,e^{Q/2}\,,
\label{hdef}
\eeq
which is  Hermithean  with respect to the scalar product in this form.
The eigenvalues of $h$ and $H$ obviously coincide.

The operator $Q$ can be
determined  perturbatively.
In full generality we write
\beq
H=H_0 +\lambda H_1
\eeq
and in the  the perturbative expansion $Q=\lambda Q_1 + \lambda^3 Q_3 + \cdots$
we may derive the terms $Q_i$ using
 Eq. (\ref{pseudoH}) and
matching the terms at each order in $\lambda$. Here we shall restrict to the
first two orders in perturbation theory
\bea
&&\left[ H_0,\frac{Q_1}{2} \right]=-H_1 \nonumber \\
&&\left[ H_0,\frac{Q_3}{2} \right]=
- \frac{1}{3} \left[\left[ H_1,\frac{Q_1}{2} \right],\frac{Q_1}{2}\right]\, .
\eea
Inserting  this into (106) for  for $h=h^{(0)}+\lambda^2
h^{(2)}+\lambda^4 h^{(4)}+\cdots$
one then obtains
\bea
&& h^{(0)}=H_0 \nonumber \\
&& h^{(2)}=\frac{1}{2} \left[H_1,\frac{Q_1}{2}\right] \nonumber \\
&& h^{(4)}=\frac{1}{2} \left[H_1,\frac{Q_3}{2}\right]+
\frac{1}{8} \left[ \left[H_0,\frac{Q_3}{2}\right],\frac{Q_1}{2}\right] \,.
\label{hpert}
\eea

Let us now use these results for the system under investigation.
We define the operators
\beq
N=\phi^\dagger \, \phi \quad,\quad R_n=N^n \phi+\phi^\dagger N^n \quad,\quad
S_n=N^n \phi-\phi^\dagger N^n
\eeq
and write $H_0=-\mu N$ and $H_1=i R_1$. One easily finds that
\beq
\frac{Q_1}{2}=-\frac{i}{\mu} S_1 \quad , \quad
\frac{Q_3}{2}=\frac{4 i}{\mu^3}\left( S_2 + \frac{1}{3} S_1 \right)
\eeq
Inserting  this expressions into Eq.~(\ref{hpert}) one finally obtains
for $h$ up
to order $\lambda^4$:
\beq
h= - \mu \left[ N + \frac{\lambda^2}{\mu^2} \left( 3N^2 -N\right)
+ \frac{\lambda^4}{\mu^4}\left(-12
  N^3+6N^2-2N+\Delta h^{(4)}\right) +{\cal O}
  \left(\frac{\lambda^6}{\mu^6}\right)\right]\,,
\label{Epert}
\eeq
where
\beq
\Delta h^{(4)}=\frac{5}{2}\left(N(N+1)\phi^2+\phi^{\dagger 2}
  N(N+1)\right)
\eeq
is a term not diagonal in the number operator $N$ and thus contributes to the
energy eigenvalues only at a higher order in $\lambda/\mu$.

Note that the transformation (\ref{iengoform}) for positive $u$ and $\lambda$
leads to Hamiltonian (\ref{Hiengoform}), which has all eigenvalues positive.
In contrast the perturbative eigenvalues are seemingly non-positive
for small enough $\lambda$ and certainly such at $\lambda=0$.
This shows  that the perturbative series is not   convergent.
Note however that it provides a reasonable asymptotic expansion
at small values of $\lambda$ for not very high rapidities and
describes well the initial period of evolution when the amplitudes rise.
As an illustration, using this perturbatively approximated spectrum  one may compute
the  pomeron propagator at two loops using
\beq
G(y)=\langle 0|  \phi\, e^{-y H}\, \phi^\dagger |0\rangle=
\langle 0|  \phi\, e^{Q/2}e^{-y h} e^{-Q/2}\, \phi^\dagger |0\rangle \,,
\eeq
expanded in the basis $|n \rangle_h$ of the eigenstates of $h$ ($h |n\rangle_h = E_n
|n\rangle_h$).
Comparison to the exact numerical evolution in the next section
shows a perfect agreement up to some
value of $\mu y$, much below the saturation region, beyond which the
perturbative approximation, which continues to grow in rapidity,
fails. In particular for $\lambda/\mu=1/10$ the perturbative approximation is
very good up to $\mu \,y\le 3$.

In the limit $\lambda\to 0$ one finds that the initial period of evolution, when
the amplitude grows, extends to infinitely high rapidities
and the propagator continuosly passes into the free one, growing as $\exp(\mu y)$.
So in spite of the essential singularity at $\lambda=0$,  the amplitudes seem
to have a well defined limit as $\lambda\to 0$, when they
continuously pass into free ones.

One might expect reasonable to apply a similar perturbative approach for some
interval of rapidity also in the case for non zero transverse dimensions.
\section{Numerical calculations}
As mentioned in the introduction,
most of the qualitative results in the toy model were obtained thirty
years ago ~\cite{amati1,alessandrini, jengo,amati2,ciafaloni1,ciafaloni2}.
They were restricted to the case of very
small triple pomeron coupling $\lambda$, when the dynamics becomes
especially transparent and admits asymptotic estimates. However
the smallness of $\lambda$ in the zero-dimensional case actually does not
correspond to the physical situation in the world with two transverse
dimensions. If the transversce space is approximated by a
two-dimensional lattice with intersite distance $a$ then
the effective coupling constant in the zero-dimensional
world is inversely proportional to $a$ and thus goes to infinity in the
continuum space (see ~\cite{amati2}). So the behaviour of the model at
any values of $\mu$ and
$\lambda$ has a certain interest. In particular it may be interesting
to know the behaviour at $\mu\to 0$, since the functional integral
defining the
theory for $\mu<0$ diverges at $\mu>0$ and one can expect a certain
singularity at $\mu=0$. The validity of different approximate methods to
find the solution
may also be of interest in view of applications to more realistic models.

Modern calculational facilities allow to find solutions for the model
at any values of $\mu$ and $\lambda$ without any difficulty.
Our approach was to evolve the wave function in rapidity directly from
its initial value at $y=0$ using Eq. (\ref{evol1}) by the Runge-Cutta technique.
This straightforward approach proved to be simple and powerful enough to
give reliable results for a very short time. True, the step in rapidity
had
to be chosen rather small and correlated with the step in $u$.
In practice we chose the initial interval of $u$ as
\beq
0<u<20
\eeq
divided in 2000 points. The corresponding step in rapidity had to be
taken not greater than $2.5\cdot 10^{-3}$

Presenting our results we have in mind that amplitudes depend
not only on the dynamics but also on the two coupling constants
$g_{i(f)}$.
To economize we therefore restrict ourselves to show either
the full pomeron propagator
\beq
P(y)=-\langle ve^{-Hy}u\rangle
\eeq
or the amplitude for the process which mimics hadron-nucleus scattering
\beq
A(y,g)=-\langle e^{-gv}e^{-Hy}u\rangle
\eeq
(with $g_f\equiv g>0$ and $g_i=1$).

\subsection{Pomeron propagator}
Straightforward evolution according to Eq. (\ref{evol1}) from the initial
wave function $F_i(0,u)=u$ gives the results for the pomeron propagator
shown in Fig. 1 for different sets of $\mu$ and $\lambda$.
We recall that the free propagator is just
\beq
P_0(y)=e^{\mu y}.
\label{freepom}
\eeq
Inspection of curves in Fig. 1 allows to make the following conclusions.

1) Inclusion of the triple-pomeron interaction of any strength
makes the propagator  fall with rapidity. This fall is the stronger
the larger the coupling.

2) At very small values of the coupling ($\rho\sim 10$ or greater)
the fall is not felt at maximal rapidity $y=50$ chosen for the evolution.
This is in full accord with the predictions of earlier studies
~\cite{jengo,amati2,ciafaloni1},
in which it
was concluded that the behaviour should be $\sim \exp(-\Delta E y)$
with $\Delta E$ given by (\ref{DeltaE}).
With $\mu=1$ and  $\rho=10$ one finds $\Delta E \sim 10^{-21}$, so that the fall
of the propagator should be felt only at fantastically large rapidities.

3) However already with $\mu=1$ and $\lambda=1/3$ the decrease of the
propagator with the growth of $y$ is  visible (see the non-logarithmic
plot in Fig. 3).

4) With small values of $\mu$ at $\lambda=1$ the propagator goes down
practically as $\exp(-y)$, so that the triple pomeron coupling takes the
role of the damping mechanism

5) No singularity is visible  at $\mu=0$ and fixed
$\lambda=1$.
The curves with very small $\mu=\pm 10^{-4}$ look completely identical.
Of course this does not exclude discontinuities in higher derivatives,
which we do not see numerically.

We next compared our exact results with two approximate estimates.
The first is the propagator obtained in the approximation of
'big loops', Eq. (\ref{pomBLsol}). The second is the asymptotic expression at small
$\lambda$ obtained in earlier studies
\beq
P(y)=\rho^2e^{- y\Delta E}
\label{asymptpom}
\eeq
with $\Delta E$ given by (\ref{DeltaE}). The results for $\mu=1$ and $\lambda
=1/10$
and 1/3 are shown in Figs. 2 and 3 respectively. As one observes, both
approximations are not quite satisfactory.
For quite small $\lambda=1/10$ the asymptotic limit, which we followed up to
 $y\sim 50$, is
correctly reproduced
in both approximations. However the exact propagator seems to reach this
limit at still higher values of $y$ (to start falling at fantastically
large $y\sim 10^{20}$). The behaviour at small values of $y$ is of course
not described by (\ref{asymptpom}) at all. The 'big loop' approximation works better
and describes the rising part of the curve with an error of less than 10\%,
which goes down to 2\% for $y \ge 15$.
At larger values of $\lambda$ the situation becomes worse. As seen from
Fig. 3 ($\lambda = 1/3$) both approximations work very poorly.
The 'big loop' approximation fails to reproduce both the magnitude of
the propagator and its  fall  at large $y$. It describes the propagator
more or less satisfactorily only at  small values of $y\leq 2$.
The approximation (\ref{asymptpom}) describes the trend of the propagator
at higher $y$ better but also fails to reproduce its magnitude.
For still higher values of $\lambda$ both approximations do not work at
all.
So our conclusion is that both approximations are applicable only at quite
small values of $\lambda/\mu<1/5$ and that the approximation of 'big loop' is
better, since it also reproduces the  growth of the propagator
at smaller $y$ from its initial value $P(0)=1$ to its asymptotic value
$\rho^2$ at maximal rapidities considered.

In Fig. 4 we compare the exact pomeron propagator with the calculations based on the
perturbative expansion (up to two loops) and with the free propagator (\ref{freepom}).
One observes that there exists a region of intermediate values of $\mu y$,
from 1 to approximately 3 where, on the one hand, the influence of loops is already
noticeable and, on the other hand, the perturbative approach works with a reasonable
precision.

\subsection{hA amplitude}
In Fig. 5 we show our results for the amplitude $A(y,g)$ at
comparatively large rapidity $y=10$ as a function of coupling $g$ to the
target for different sets of $\mu$ and $\lambda$. This dependence mimics
that of the realistic hA amplitude on the atomic number of the nucleus.
In all cases except for $(\mu,\lambda)=(-1,1/10)$ the amplitude $A(g)$
rapidly grows  as
$g$ rises from zero to approximately unity and  then  continues to
grow very
slowly at  higher values of $g$ clearly showing signs of saturation. To
make this
behaviour visble we use the log-log plot. In the exceptional case
$(\mu,\lambda)=(-1,1/10)$ one expects a similar saturation but at higher
values of $g$, as will become clear in the following.

We again compare our exact results with predictions of pure fan
approximation, Eq.(\ref{fansolhA}) with $g_i=1$ and $u=g$, and the asymptotic
formula similar to Eq. (\ref{asymptpom})
\beq
A(y,g)=\rho\Big(1-e^{-g\rho}\Big)e^{-y\Delta E}.
\eeq
The results are
shown in Figs. 6,7 and 8 for
$(\mu,\lambda)=(1,1/10), (1,1/3)$ and (-1,1/10) respectively.
With $\mu$ positive the effect of loops is considerable, so that pure fan
model gives a bad description, especially at low values of $g$ and not very
small values of $\lambda$.
The asymptotic formula  (120) works much better. At $\mu=1$ and
$\lambda=1/10$ it gives very good results, as seen from Fig. 6. However at
larger values of $\lambda/\mu$ its precision rapidly goes down.
The case $\mu=-1,\lambda=1/10$ illustrated in Fig. 8 was chosen to see the
influence of loops for the subcritical pomeron, where such influence
should be minimal. In this case the pure fan formula  gives
the amplitude
\beq
A(y=10,g)=\frac{ge^{-y}}{1+10g(1-e^{-y})}\Big|_{y=10}
\eeq
which grows linearly with $g$ up to values around $g=10$ and only
then saturates at the value $0.1e^{-10}$. This explains a somewhat
exceptional behaviour of $A(y=10,g)$ for such $\mu$ and $\lambda$.
As follows from Fig. 8 the influence of loops is in fact small. However
it is greater than $\sim\lambda^2$=1\%  as one could expect on simple
estimates  of a single loop contribution. In fact the pure fan formula
overshoots the exact result by about 20\%.

\section{Conclusions}
As mentioned, the zero-dimensional RFT was studied in detail some thirty
years ago. Our aim was to clarify some mathematical aspects of the model and also to
study it at different values of the triple pomeron coupling and not only at small ones.

We have found that the model can be consistently formulated in terms of the
quantum Hamiltonian. This formulation is valid both for negative and positive values
of $\mu$, in contrast to the functional approach which does not admit positive $\mu$.
The Hamiltonial formulation does not need to introduce a scalar product in
the representation in which the creation operators $u$ are diagonal. The wave functions
seem to be  not integrable in the whole complex $u$ plane. Rather the standard scalar
product should be used in which the creation and annihilation operators are expressed
by Hermithean operators.

A clear physical content of the theory is well seen after transformation to an
Hermithean Hamiltonian made in ~\cite{jengo}. However such transformation can only be
done separately for positive and negative $u$, leading to two branches of the spectrum
for the
initial Hamiltonian, which include both positive anfd negative eigenvalues.
The choice between two branches is determined by the signs of $\lambda$ and
external coupling constants.

For $\mu>0$ a simple case of pure fan diagrams admits an explicit
solution and reinterpretation in terms of
fashionable reaction-diffusion approach. Using this one can obtain an approximate
expression for the pomeron propagator with loops of the largest extension in rapidity
taken ito account, which gives a reasonable approximation at small values of $\lambda$.

Finally we performed numerical calculations of the  amplitudes.
They show that in all cases the triple pomeron interaction makes amplitudes fall
at high rapidities. This fall starts later for smaller $\lambda$ and at very
small $\lambda$ begins at asymptotically high rapidities (for $\lambda/\mu<1/4$
it is noticeable only at $\mu y>100$). At small $\lambda$ the behaviour in $y$ is well
described by formulas obtained in earlier studies.
However when $\lambda$ is not so small all approximate formulas work poorly.
Our numerical calculations have also shown that there is no visible singularity at
$\mu=0$, in spite of the fact that the functional formulation meets with
a divergence at this point.

We have proposed a new method for the perturbative treatment of the theory,
which may have appliciations to more sophisticated models in the world of more
dimensions

\section{Acknowledgments}
M.A.B. and G.P.V. gratefully acknowledge the hospitality of the II
Institut for Theoretical Physics of the University of Hamburg, where part of
this work was done. M.A.B. also thanks  the Bologna Physics department and
INFN for hospitality. The authors thank J. Bartels,
S.Bondarenko, L.Motyka and A.H. Mueller for very interesting and constructive
discussions.
G.P.V. thanks the Alexander Von Humboldt foundation for partial support.
This work was also partially supported by grants RNP 2.1.1.1112 and RFFI 06-02-16115a
of Russia.

%
%
\begin{figure}[ht]
\epsfxsize 4in
\centerline{\epsfbox{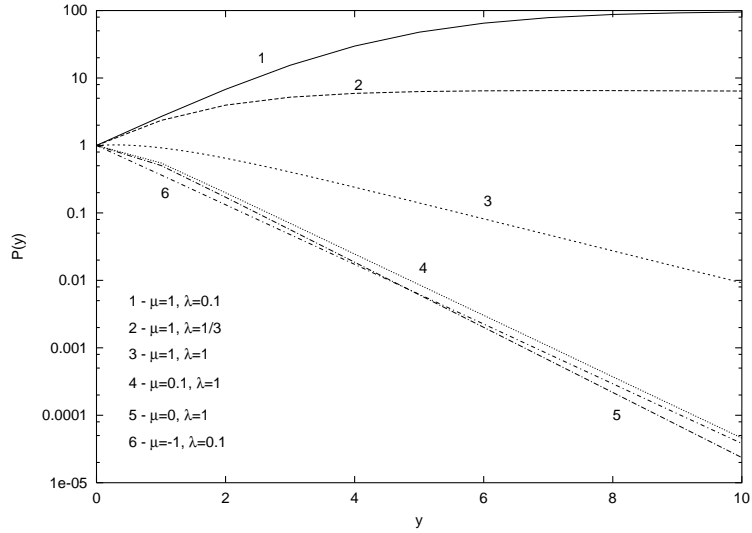}}
\caption{The full pomeron propagator as a function of rapidity for
different sets of $\mu$ and $\lambda$ indicated in the figure}
\label{Fig1}
\end{figure}
%
\begin{figure}[ht]
\epsfxsize 4in
\centerline{\epsfbox{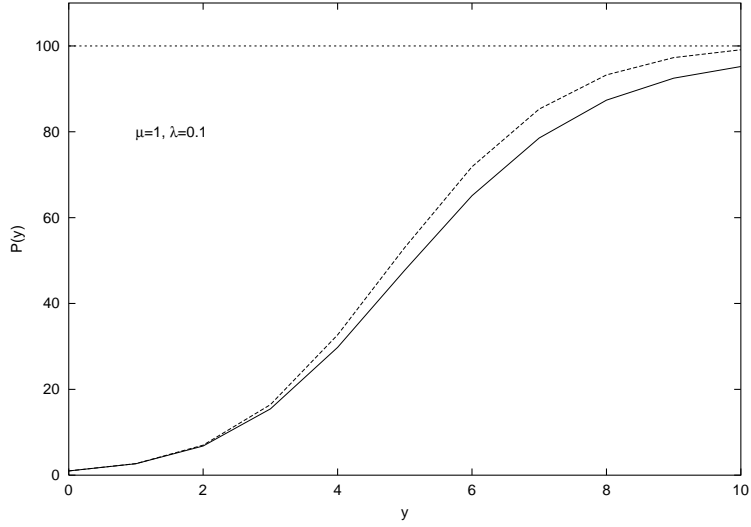}}
\caption{The full pomeron propagator as a function of rapidity
for $\mu=1$ and $\lambda=0.1$
(the lower curve) as compared to the 'big loop' formula Eq. (88)
(the middle curve) and asymptotic expression (119) (the upper curve)}
\label{Fig2}
\end{figure}
%
\begin{figure}[ht]
\epsfxsize 4in
\centerline{\epsfbox{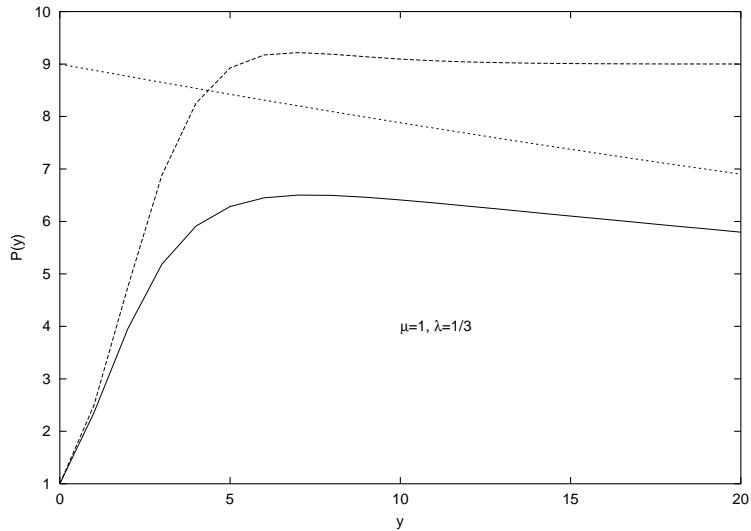}}
\caption {The full pomeron propagator as a function of rapidity
for $\mu=1$ and $\lambda=1/3$
(the lower curve) as compared to the 'big loop' formula Eq. (88)
(the upper curve on the right) and asymptotic expression
(119) (the middle curve on the right)}
\label{Fig3}
\end{figure}
%
\begin{figure}[ht]
\epsfxsize 4in
\centerline{\epsfbox{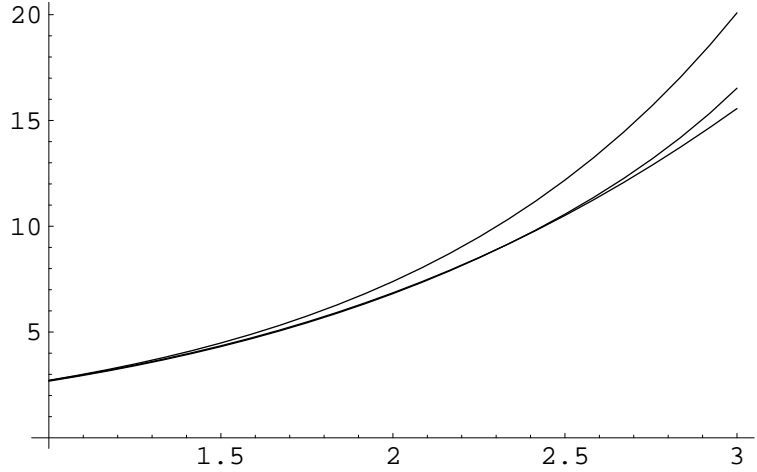}}
\caption{The Pomeron Green's function as a function of $\mu\, y$ for $\mu=1$ and $\lambda=1/10$: from
  bottom to top: the exact numerical evolution,  the two loop
  perturbative approximation, the free case ($e^{y \mu}$).  }
\label{Fig4}
\end{figure}
%
\begin{figure}[ht]
\epsfxsize 4in
\centerline{\epsfbox{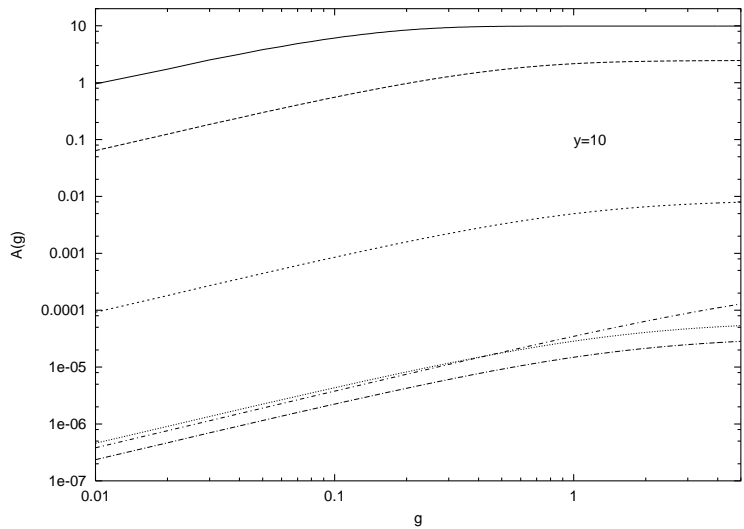}}
\caption{ The hA amplitude at $y=10$ as a function of the coupling
to the target. Curves from top to bottom on the right correspond to
$(\mu,\lambda)$=(1,0.1),(1,1/3), (1,1),(-1,0.1),(0.1,1),(0,1)}
\label{Fig5}
\end{figure}
%
\begin{figure}[ht]
\epsfxsize 4in
\centerline{\epsfbox{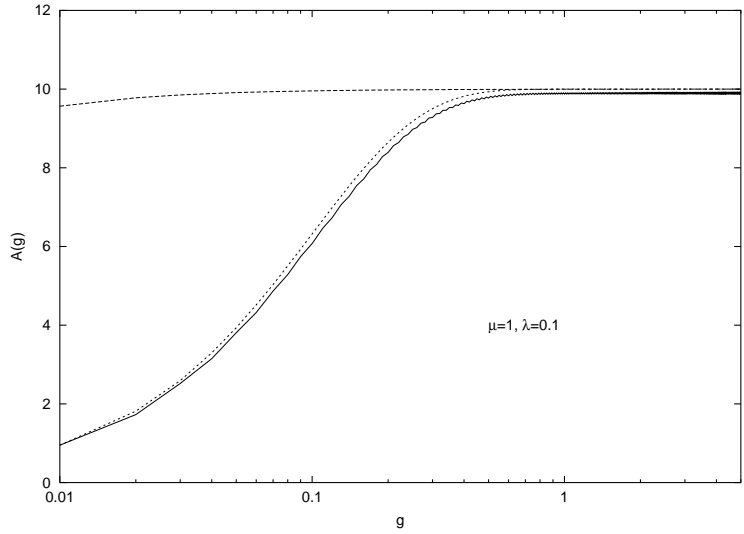}}
\caption{the hA amplitude $A(y,g)$ at $y=10$ for $\mu=1$, $\lambda=0.1$
(the lower curve) as compared to the pure fan prediction (the upper
curve) and the asymptotic expression Eq. (120) (middle curve)}
\label{Fig6}
\end{figure}
%
\begin{figure}[ht]
\epsfxsize 4in
\centerline{\epsfbox{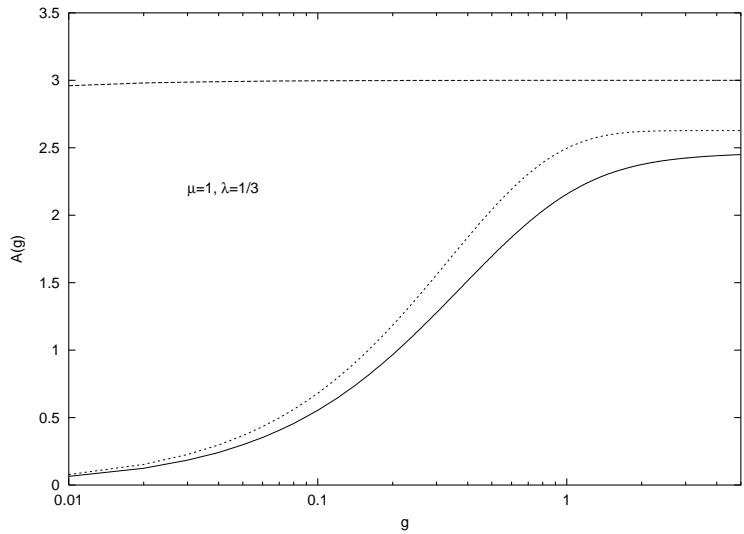}}
\caption{Same as Fig. 5 for $\mu=1$ and $\lambda=1/3$}
\label{Fig7}
\end{figure}
%
\begin{figure}[ht]
\epsfxsize 4in
\centerline{\epsfbox{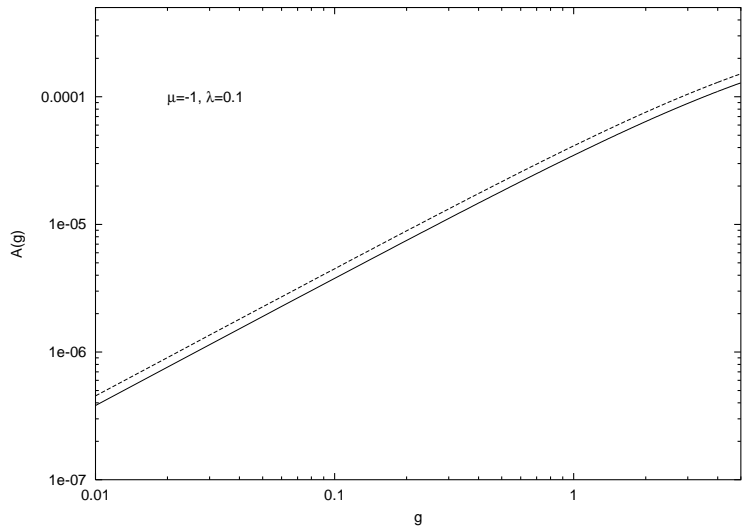}}
\caption{The hA amplitude $A(y,g) $at $y=10$ for $\mu=-1$, $\lambda=0.1$
(the lower curve) as compared to the pure fan prediction (the upper
curve)}
\label{Fig8}
\end{figure}

\end{document}